\documentclass{emulateapj}
\usepackage{gensymb,amsmath}
\usepackage{epstopdf}
\usepackage{url,hyperref}
\pdfoutput=1

\shorttitle{\mbox{CO(1--0)} in J14011}
\shortauthors{Sharon et al.}
\slugcomment{ApJ Accepted 12/24/2012}
\begin{document}
\title{VLA Mapping of the \mbox{CO(1--0)} Line in SMM J14011+0252}

\author{Chelsea E. Sharon, Andrew J. Baker}
\email{csharon@physics.rutgers.edu}
\email{ajbaker@physics.rutgers.edu}
\affil{Department of Physics and Astronomy, Rutgers, the State University of New Jersey, Piscataway, NJ, 08854-8019}
\author{Andrew I. Harris}
\email{harris@astro.umd.edu}
\affil{Department of Astronomy, University of Maryland, College Park, MD 20742}
\and
\author{Alasdair P. Thomson}
\email{at@roe.ac.uk}
\affil{Institute for Astronomy, University of Edinburgh, Blackford Hill, Edinburgh EH9 3HJ, UK}

\begin{abstract} 
We present high-resolution \mbox{CO(1--0)} observations of the lensed submillimeter galaxy (SMG) SMM\,J14011+0252 at $z=2.6$. Comparison to the previously-detected \mbox{CO(3--2)} line gives an intensity ratio of $r_{3,1}=0.97\pm0.16$ in temperature units, larger than is typical for SMGs but within the range seen in the low-$z$ ultraluminous infrared galaxy population. Combining our new data with previous mid-$J$ CO observations, we perform a single-phase large velocity gradient (LVG) analysis to constrain the physical conditions of the molecular gas. Acceptable models have significant degeneracies between parameters, even when we rule out all models that produce optically thin emission, but we find that the bulk of the molecular gas has $T_{\rm kin}=20$--$60\,{\rm K}$, $n_{\rm H_2}\sim10^4$--$10^5\,{\rm cm^{-3}}$, and $N_{\rm CO}/\Delta v=10^{17.00\pm0.25}\,{\rm cm^{-2}\,km^{-1}\,s}$. For our best-fit models to self-consistently recover a typical CO-to-${\rm H_2}$ abundance and a plausible degree of virialization, the local velocity gradient in the molecular gas must be substantially larger than its galaxy-wide average. This conclusion is consistent with a scenario in which SMM\,J14011+0252 has a fairly face-on orientation and a molecular ISM composed of many unresolved clouds. Using previous ${\rm H\alpha}$ observations, we find that SMM\,J14011+0252 has a spatially resolved star formation rate vs.~molecular gas surface density relation inconsistent with those of ``normal" local star-forming galaxies, even if we adopt a local ``disk-like" CO-to-${\rm H_2}$ conversion factor as motivated by our LVG analysis. This discrepancy supports the inference of a star formation relation for high-$z$ starbursts distinct from the local relation that is not solely due to differing choices of gas mass conversion factor.

\end{abstract}


\section{Introduction \label{sec:intro}}

Since the discovery of a population of dusty, high-redshift submillimeter galaxies (SMGs), whose high star formation rates (SFRs) mean they contribute substantially to the cosmic infrared background (see review article by \citealt{blain2002}), significant efforts have been made to characterize their star-forming molecular gas \citep[e.\/g.\/,][]{frayer1998,frayer1999,neri2003,greve2005,tacconi2006,tacconi2008}. While many of these studies have used solely mid-$J$ CO lines (i.\/e.\/, \mbox{CO(3--2)}, \mbox{CO(4--3)}, and \mbox{CO(5--4)}), recent expansions of the frequency coverage of radio telescopes into the Ka band have allowed observations of \mbox{CO(1--0)}---the best tracer of cold and low-density molecular gas---for SMGs whose redshifts fall near the $z\sim2$--$3$ peak of the distribution for radio-identified samples \citep{chapman2005}. These build on observations of \mbox{CO(1--0)} for other $z>1$ poulations in other bands (e.\/g.\/, \citealt{papadopoulos2001,carilli2002,greve2003,greve2004,klamer2005,hainline2006,riechers2006,dannerbauer2009,carilli2010,aravena2010}). The SMG studies have shown that the \mbox{CO(3--2)} to \mbox{CO(1--0)} line ratio favors a value of $r_{3,1}\sim0.6$ (in units of brightness temperature), indicating the presence of a multi-phase molecular ISM that includes a substantial cold gas reservoir \citep{swinbank2010,harris2010,ivison2011,danielson2011}. While similar line ratios have been observed in other types of high-redshift star-forming systems like Lyman break galaxies \citep{riechers2010b} and BzK-selected galaxies \citep{dannerbauer2009,aravena2010}, and in low-redshift dusty galaxies (albeit with a much wider range of $r_{3,1}$; e.\/g.\/, \citealt{mauersberger1999,yao2003}), this result is in direct contrast to those for quasar host galaxies, whose near-unity value of $r_{3,1}$ is consistent with single-phase molecular gas (\citealt{riechers2006,riechers2011f}; see also $r_{4,1}\gtrsim1$ in \citealt{ivison2012}). 

While these \mbox{CO(1--0)} measurements have been essential for determining the {\it global} multi-phase nature of the molecular gas in SMGs, the lack of high-resolution mapping leaves some margin for error in interpreting line ratios. In order to be confident in assessing physical conditions in the gas, we need to know that the CO lines are being emitted from the same localized regions. Given the clumpy structure observed in SMGs (e.\/g.\/, \citealt{tacconi2008,swinbank2010}), and the role major mergers are likely playing in the star formation bursts of (at least) some of these systems (e.\/g.\/, \citealt{conselice2003,narayanan2010,engel2010,dave2010}), it is not safe to assume a homogenous distribution of gas conditions within every galaxy \citep[e.\/g.\/,][]{mao2000,zhu2003}. In addition, since many of the well-studied SMGs are gravitationally lensed, it is possible that differential lensing (the variation in magnification factor across a spatially extended source) could be affecting the observed line ratios \citep[e.\/g.\/,][]{blain1999,serjeant2012}. Indirect evidence for the spatial variation of excitation in SMGs has been seen in the more extended distribution of \mbox{CO(1--0)} relative to radio continuum emission (from the synchrotron radiation from supernova remnants, which trace more actively star-forming and thus higher-excitation gas; \citealt{ivison2011}). The larger spatial extent of SMGs' cold gas reservoirs is also bolstered by the larger line widths of \mbox{CO(1--0)} vs.~\mbox{CO(3--2)} \citep{ivison2011,thomson2012}.

The high-resolution CO observations necessary for evaluating spatial variation of CO excitation also enable {\it spatially resolved} comparisons between the Schmidt-Kennicutt relations \citep{schmidt1959,kennicutt1998} seen at low and high redshifts. Determining galaxies' gas mass surface densities from CO surface brightness requires the assumption of a CO-to-${\rm H_2}$ conversion factor. Observations of high, and sometimes unphysical, gas mass fractions in low-$z$ IR-bright starbursts and some disk galaxy centers \citep[e.\/g.\/,][]{scoville1991,garcia1993,downes1993,solomon1997,scoville1997,oka1998,hinz2006,meier2010} imply a range of  $\alpha_{\rm CO}$, leading to the adoption of a ``disk" value \citep[$\alpha_{\rm CO}=4.6\,M_\sun\,{\rm (K\, km\, s^{-1}\,pc^2)}^{-1}$;][]{solomon1991} for normal star-forming galaxies and a ``starburst" value \citep[$\alpha_{\rm CO}=0.8\,M_\sun\,{\rm (K\, km\, s^{-1}\,pc^2)}^{-1}$;][]{downes1998} for mergers and other high surface density environments. Some effort has been made to find a suitable continuous form of $\alpha_{\rm CO}$ \citep[e.\/g.\/,][]{glover2011,narayanan2012} that also matches observed trends in metallicity \citep[e.\/g.\/,][]{wilson1995,arimoto1996,israel1997,leroy2006,bolatto2008,genzel2012}. Which value (or values) of $\alpha_{\rm CO}$ is appropriate for high-redshift galaxies is debated \citep[e.\/g.\/,][]{tacconi2008}, and the apparent bimodality in the high-$z$ Schmidt-Kennicutt relation (where SMGs fall along the low-$z$ starburst track, and more ``normal" high-$z$ galaxies extend the low-$z$ disk galaxy track to higher masses and star formation rates; e.\/g.\/, \citealt{daddi2010,genzel2010}; cf. \citealt{ivison2011}) may be caused in part by the choice of $\alpha_{\rm CO}$. Comparisons between the Schmidt-Kennicutt relations at high and low redshifts may also be affected by the differences in the CO lines observed \citep{narayanan2011}, as the molecular gas at low redshift is measured via the \mbox{CO(1--0)} line, while the molecular gas at high redshift has typically been measured via mid-$J$ CO lines. While galaxy-wide measurements of the star formation rate and molecular gas mass are typical for high-$z$ systems, evaluation of the true surface density version of the Schmidt-Kennicutt relation \citep[e.\/g.\/,][]{kennicutt2007,bigiel2008,wei2010} requires spatially resolved maps of CO lines and star formation tracers.

In order to explore the spatial variation of CO line ratios and the Schmidt-Kennicutt relation for high-$z$ starbursts, we have obtained high-resolution Karl G. Jansky Very Large Array (VLA) mapping of the \mbox{CO(1--0)} line in SMM J14011+0252 (J14011 hereafter), which we compare to previous mid-$J$ and rest-frame mid-IR observations. J14011 ($z_{\rm opt}=2.55$; \citealt{barger1999}) was discovered as part of the SCUBA Lens Survey \citep{smail1998,smail2002}; it consists of two optical/near-IR components in the source plane (J1 and J2) that are weakly gravitationally lensed by the cluster A1835 ($z=0.25$) and one of its members \citep[J1c;][]{motohara2005}, with a median magnification factor of $\mu\sim3.5\pm0.5$ \citep{smail2005}. Since its discovery, J14011 has been followed up extensively with optical and near-IR observations \citep{barger1999,ivison2000,ivison2001,tecza2004,frayer2004,swinbank2004,motohara2005,nesvadba2007}, and it was only the second SMG to be successfully detected in any CO transition \citep[\mbox{CO(3--2)};][]{frayer1999}. Further \mbox{CO(3--2)} observations were made by \citet{ivison2001}, as were \mbox{CO(3--2)} and \mbox{CO(7--6)} observations by \citet{downes2003}. Strong upper limits on its X-ray flux \citep{ivison2000,fabian2000,rigby2008} and ratios of the $7.7\,{\rm \mu m}$ polycyclic aromatic hydrocarbon (PAH) feature to mid-infrared flux \citep{rigby2008} indicate that J14011 is a pure starburst, lacking a significant contribution to its dust luminosity from an active galactic nucleus (AGN). It is similar in its far-infrared luminosity ($L_{\rm FIR}\sim 4\times10^{12}\,L_\sun$; \citealt{smail2005}) and its clumpy morphology within and between optical/near-IR components \citep{ivison2000,tecza2004,swinbank2004} to low-redshift ultra/luminous infrared galaxies (U/LIRGs) such as \mbox{Arp 220} \citep{frayer1999,ivison2000,downes2003,motohara2005,nesvadba2007,rigby2008}. Initial single-dish observations of the \mbox{CO(1--0)} line indicated $r_{3,1}=0.76\pm0.12$ \citep{harris2010}, similar to the values seen for other SMGs.

We assume the WMAP7+BAO+$H_0$ mean $\Lambda$CDM cosmology throughout this paper, with $\Omega_\Lambda=0.725$ and $H_0=70.2\,{\rm km\,s^{-1}\,Mpc^{-1}}$ \citep{komatsu2011}.

\section{Observations}
\label{sec:obs}

We observed J14011 \citep[$\alpha {\rm (J2000)} =  14^{\rm h}01^{\rm m}4.^{\rm s}93$, $\delta {\rm (J2000)} =  +02\degree52^{\prime}24.1^{\prime\prime}$; \mbox{CO(3--2)} position from][]{downes2003} at the VLA on 2010 October 1 and 3 in the DnC configuration (maximum baseline $1.868\,{\rm km}$) and 2011 January 22, 25, and 28--31 in the CnB configuration (maximum baseline $6.903\,{\rm km}$). During the DnC observations, 23 antennas were functioning and equipped with Ka-band receivers, and the weather conditions were mixed ($<70\%$ sky coverage by cumuliform clouds). During the CnB observations, 26 Ka-band-equipped antennas were available, and the weather conditions were excellent. We observed with the Wideband Interferometric Digital Architecture (WIDAR) correlator in the ``Open Shared Risk Observing (OSRO) Full Polarization" mode using the lowest available spectral resolution ($64\,{\rm channels}\times2\,{\rm MHz\, resolution}$). The intermediate frequency (IF) channel pairs A/C and B/D were tuned to overlap by ten channels, creating a full bandwidth of $236\,{\rm MHz}$ centered at $32.3703\,{\rm GHz}$ (chosen to keep the line away from IF pair edges). At the beginning of each track we observed 3C286 as a flux calibrator, adopting $S_\nu=1.9455\,{\rm Jy}$ as the Common Astronomy Software Applications (CASA) package's default ``Perley-Butler 2010" flux density\footnote{\label{casanote}\href{http://casa.nrao.edu}{http://casa.nrao.edu}}. Phase and amplitude fluctuations were tracked by alternating between the source and a nearby quasar, J1354--0206, with a cycle time of 6 minutes; this quasar was also used for passband calibration. A total of 18.7 hours was spent on source. 

Calibration and mapping were carried out in CASA\textsuperscript{\ref{casanote}}. After the tracks had been corrected for Doppler shifts relative to the local kinematic (radio) standard of rest using the CASA routine cvel, the original channels were smoothed by a factor of two, resulting in $4\,{\rm MHz}$ ($37\,{\rm km\,s^{-1}}$ rest frame) spectral resolution. The naturally weighted channel maps have a synthesized beam of $0.^{\prime\prime}68\times 0.^{\prime\prime}52$ at position angle $119.02$ degrees and an average RMS noise of $53.2\,{\rm \mu Jy\,beam^{-1}}$. Analysis of the resulting data cube was performed using a custom set of IDL scripts.

\section{Results}
\label{sec:results}

\begin{figure}
\plotone{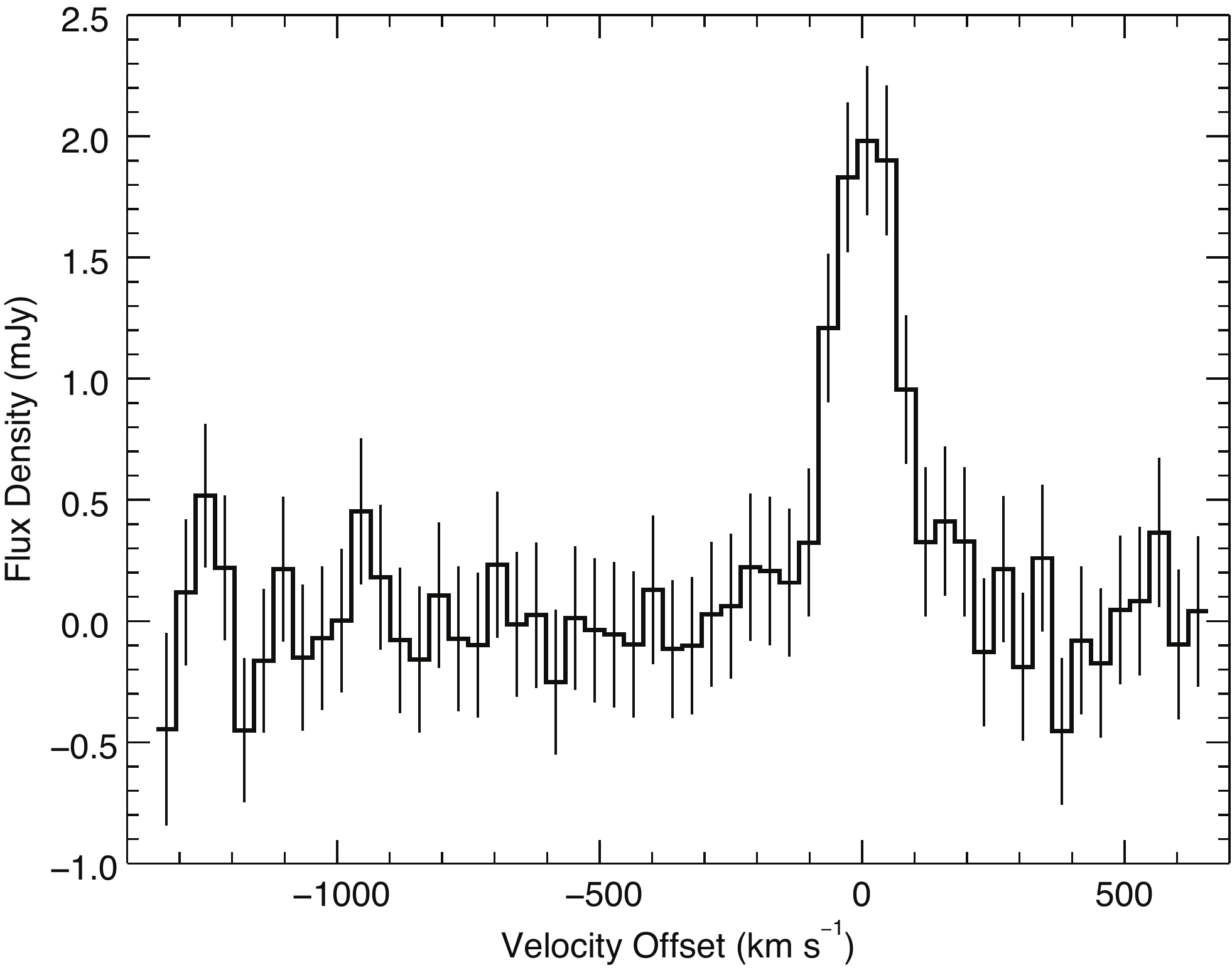}
\caption{$37\,{\rm km\,s^{-1}}$ resolution VLA spectrum plotted relative to the $z=2.5652$ \mbox{CO(1--0)} systemic redshift from the GBT/Zpectrometer observations \citep{harris2010}. Vertical lines indicate statistical uncertainties. \label{fig:spectra}}
\end{figure}

Our spectrum (Fig.\,\ref{fig:spectra}) successfully detects the \mbox{CO(1--0)} line at the $8\sigma$ level. A Gaussian fit to our line profile gives a peak flux density of $2.08\pm0.22\,{\rm mJy}$ and a FWHM of $151\pm19\,{\rm km\,s^{-1}}$. The FWHM is consistent with the \mbox{CO(7--6)} line width of $170\pm30\,{\rm km\,s^{-1}}$, and marginally narrower than the \mbox{CO(3--2)} line width of $190\pm11\,{\rm km\,s^{-1}}$ from \citet{downes2003}. The integrated \mbox{CO(1--0)} line flux (using the seven channels around the line peak) is $0.32\pm0.03(\pm0.03)\,{\rm Jy\,km\,s^{-1}}$, where the latter $10\%$ uncertainty is associated with the flux calibration. While our measured integrated line flux appears to be lower than the $\sim0.40\pm0.05\,{\rm Jy\,km\,s^{-1}}$ of \citet{harris2010}, the two values are consistent within their $1\sigma$ errors. The slight discrepancy is likely because the Zpectrometer-measured value is from a Gaussian fit in which the line width is only an upper limit (the narrow line was marginally spectrally resolved by the Zpectrometer). Convolution of the VLA spectrum with the $20\,{\rm MHz}$ FWHM sinc($x$) response function of the Zpectrometer reproduces the Gaussian fit integrated line flux given in \citet{harris2010}, indicating that our measurements are in good agreement when we account for spectral resolution differences. Using the Milky Way conversion factor $\alpha_{\rm CO}=4.6\,M_\sun\,{\rm (K\, km\, s^{-1}\,pc^2)}^{-1}$ \citep{solomon1991} suggested by our LVG analysis (Section\,\ref{sec:lvg}) and correcting for lensing magnification and helium, the molecular gas mass is $(1.9\pm0.3)\times10^{11}\,M_\sun$. The redshift as determined by our \mbox{CO(1--0)} measurement is $z=2.5653\pm0.0001$, which is consistent with the previous optical \citep[e.\/g.\/,][]{barger1999,ivison2000} and CO \citep[e.\/g.\/,][]{downes2003,harris2010} redshifts.

\begin{figure}
\plotone{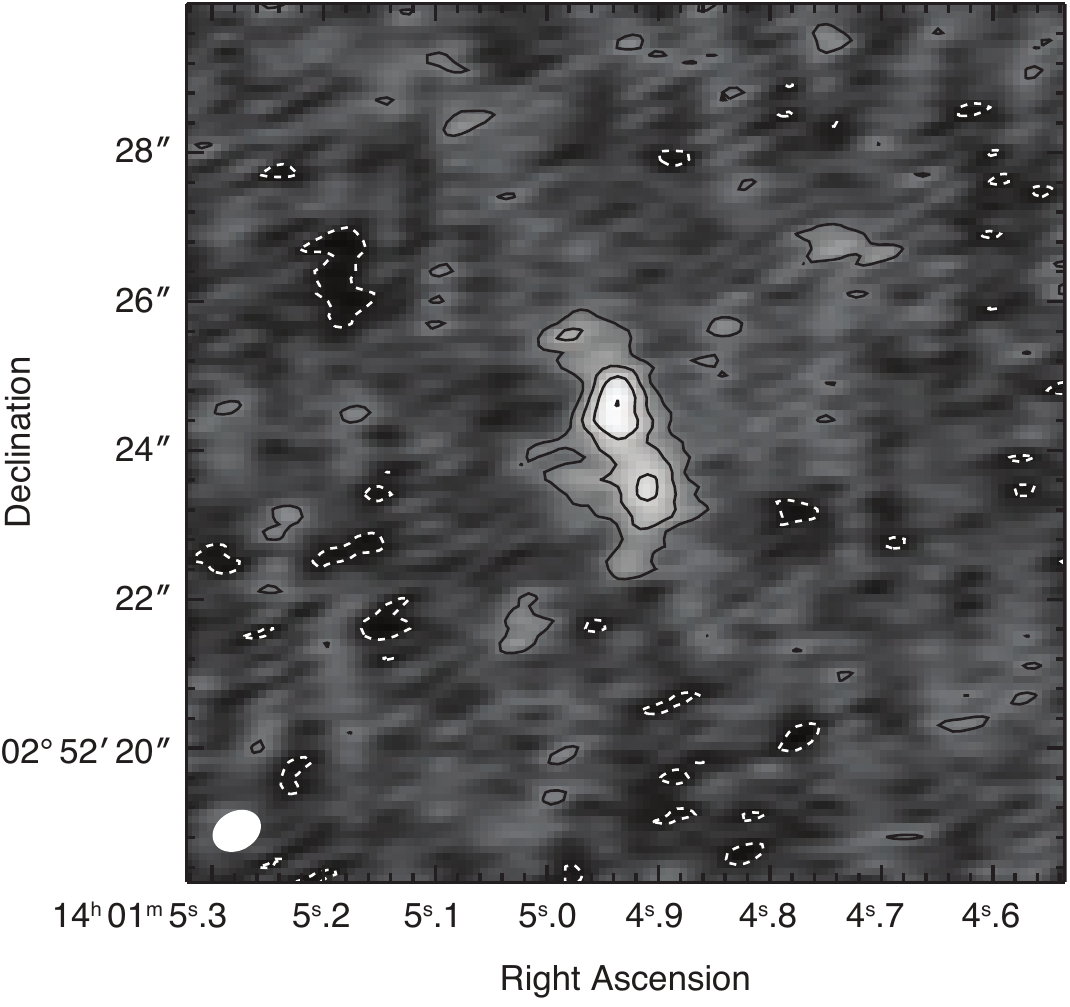}
\caption{Integrated \mbox{CO(1--0)} intensity map. Contours are multiples of $\pm2\sigma$ ($\sigma=0.16\,{\rm mJy\,beam}^{-1}$), where negative contours are dotted; synthesized beam is shown at lower left. \label{fig:intlinemap}}
\end{figure}

\begin{figure}
\plotone{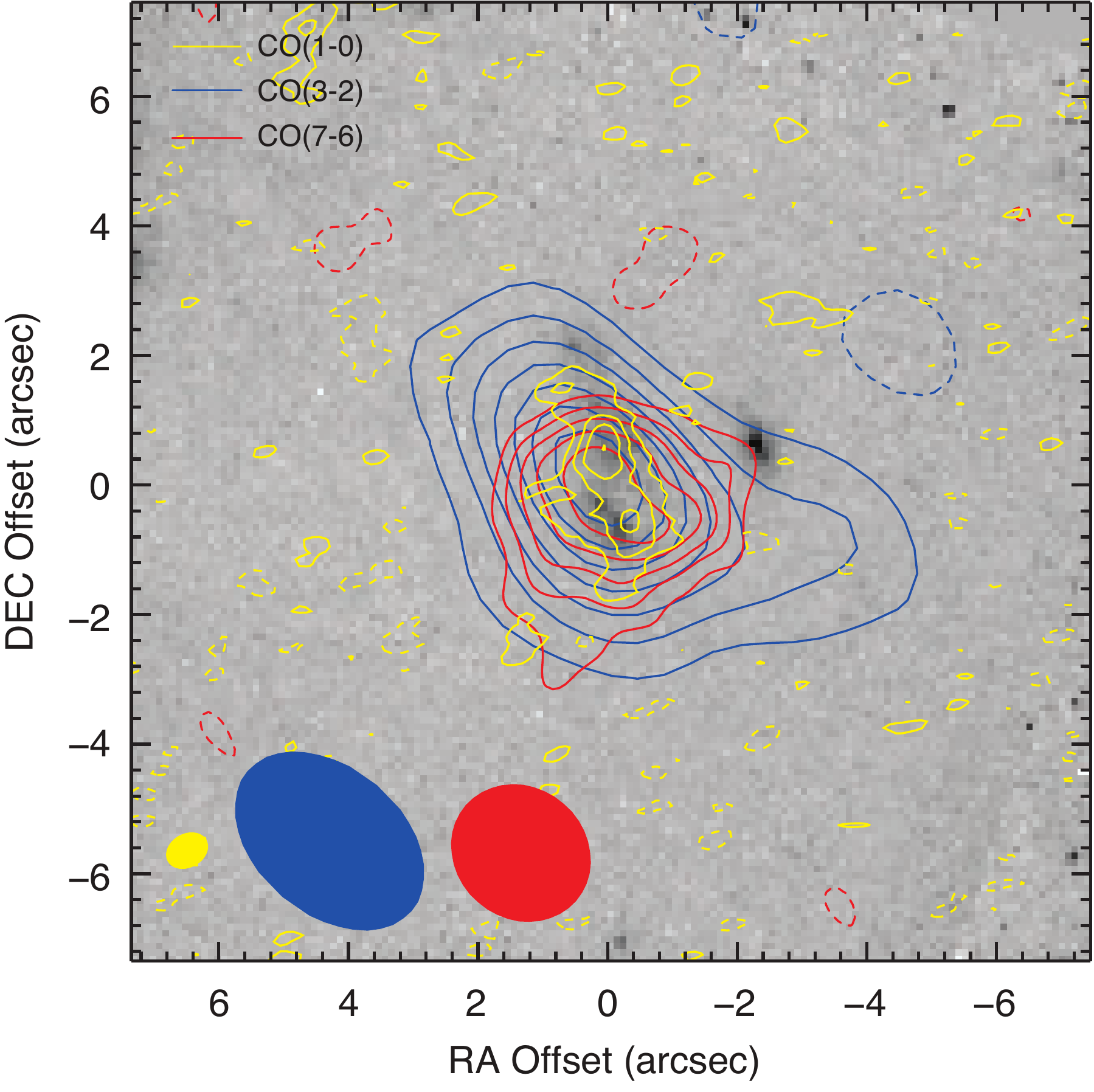}
\caption{Overlaid contours of all published CO maps of J14011 and the \citet{aguirre2012} \emph{HST} F160W image (with the foreground interloper, J1c, removed), centered at $\alpha {\rm (J2000)} =  14^{\rm h}01^{\rm m}04.^{\rm s}93$ and $\delta {\rm (J2000)} =  +02\degree52^{\prime}24.^{\prime\prime}1$. The \mbox{CO(1--0)}, CO(3--2), and CO(7--6) maps are in yellow, blue, and red respectively; the synthesized beam for each map is shown by the filled ellipse of the corresponding color in the lower left corner. The \mbox{CO(1--0)} line contours are as described in Fig.\,\ref{fig:intlinemap}. The CO(3--2) and CO(7--6) observations are taken from \citet{downes2003}. The CO(3--2) contours are multiples of $\pm2\sigma$, where $\sigma=0.41\,{\rm mJy\,beam}^{-1}$. The CO(7--6) contours are in steps of $\pm1\sigma$, starting at $\pm2\sigma$, where $\sigma=1.3\,{\rm mJy\,beam}^{-1}$. \label{fig:allmaps}}
\end{figure}

\begin{figure*}
\includegraphics[scale=0.42]{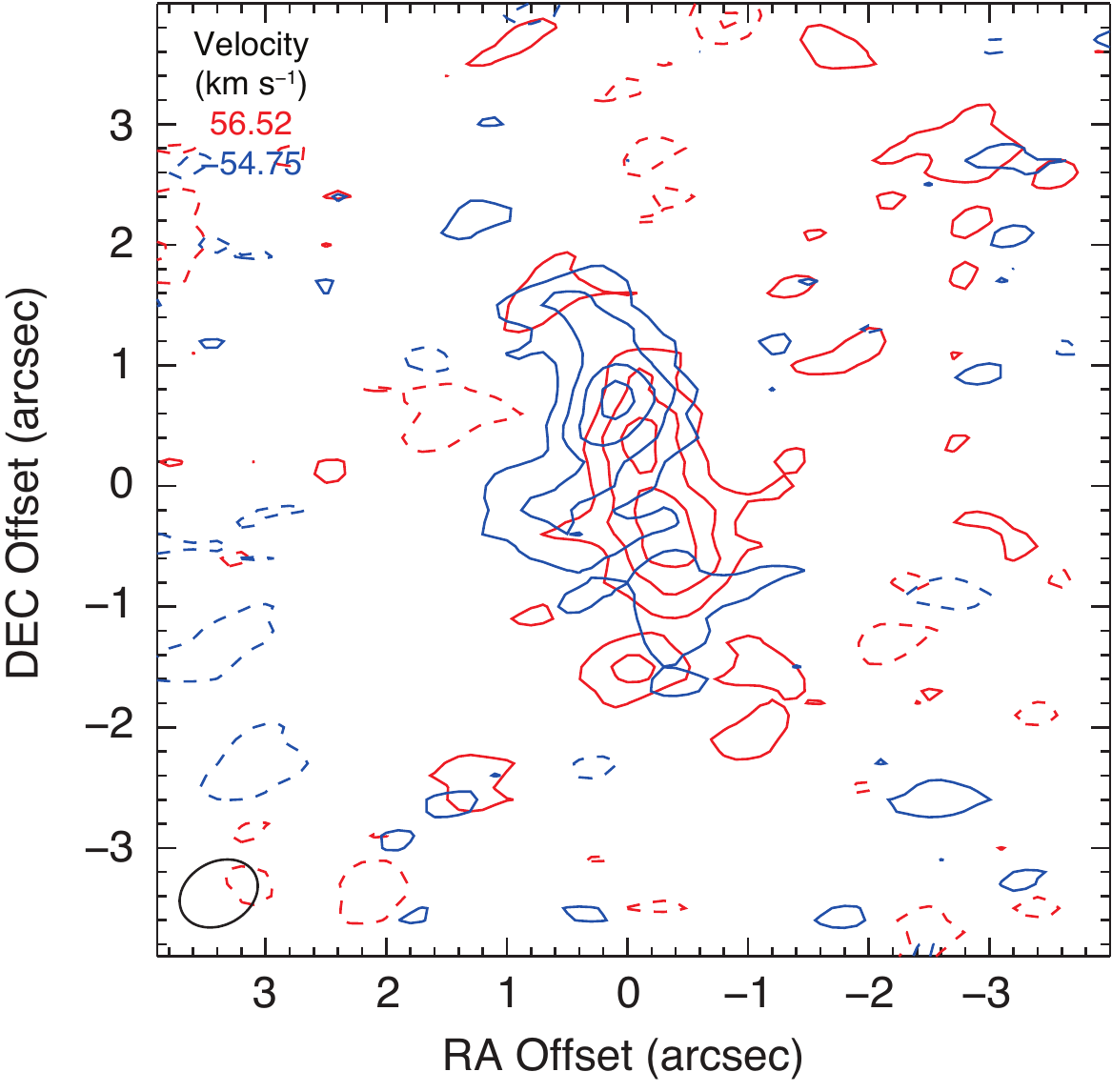}
\includegraphics[scale=0.42]{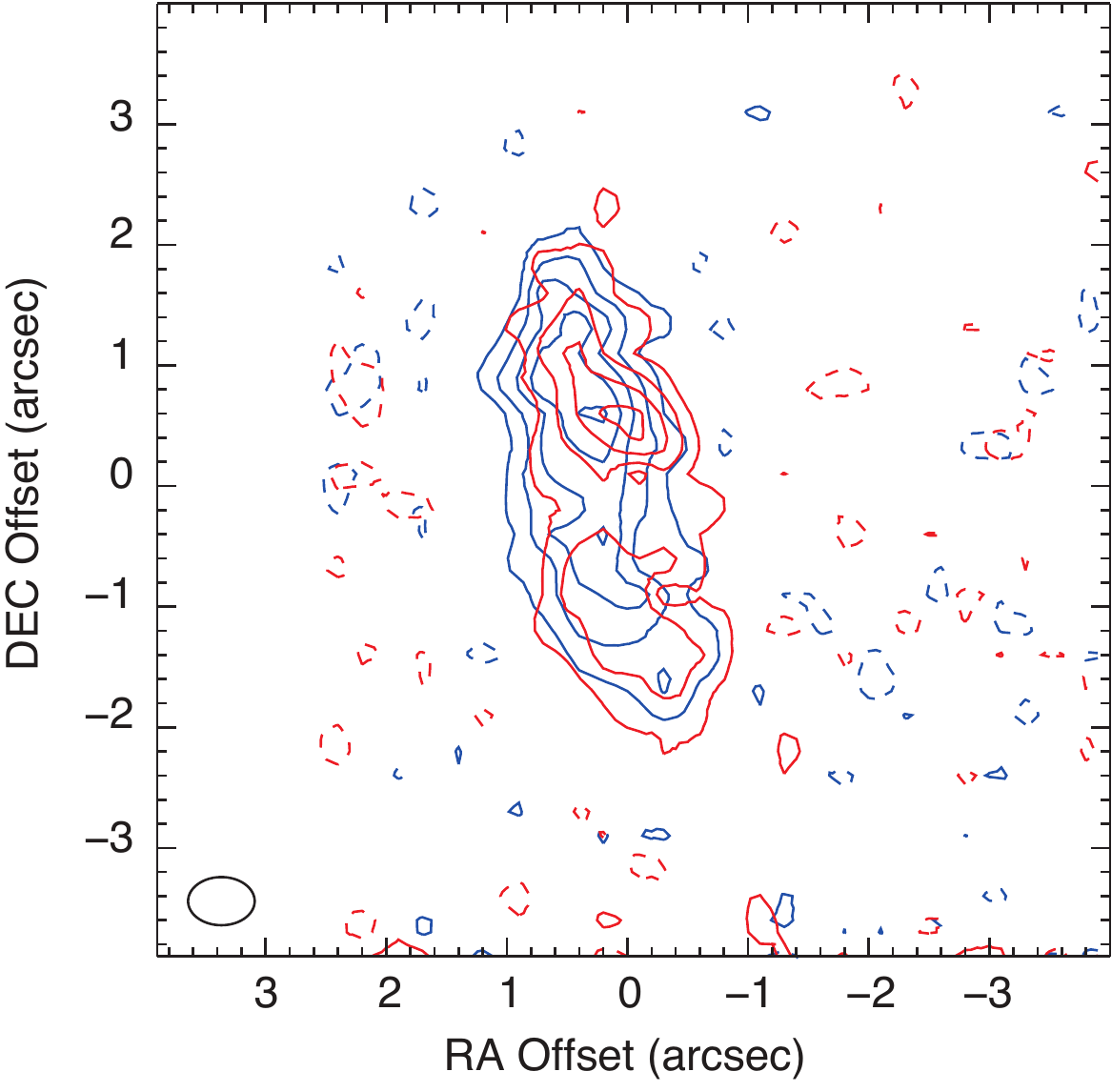}
\includegraphics[scale=0.42]{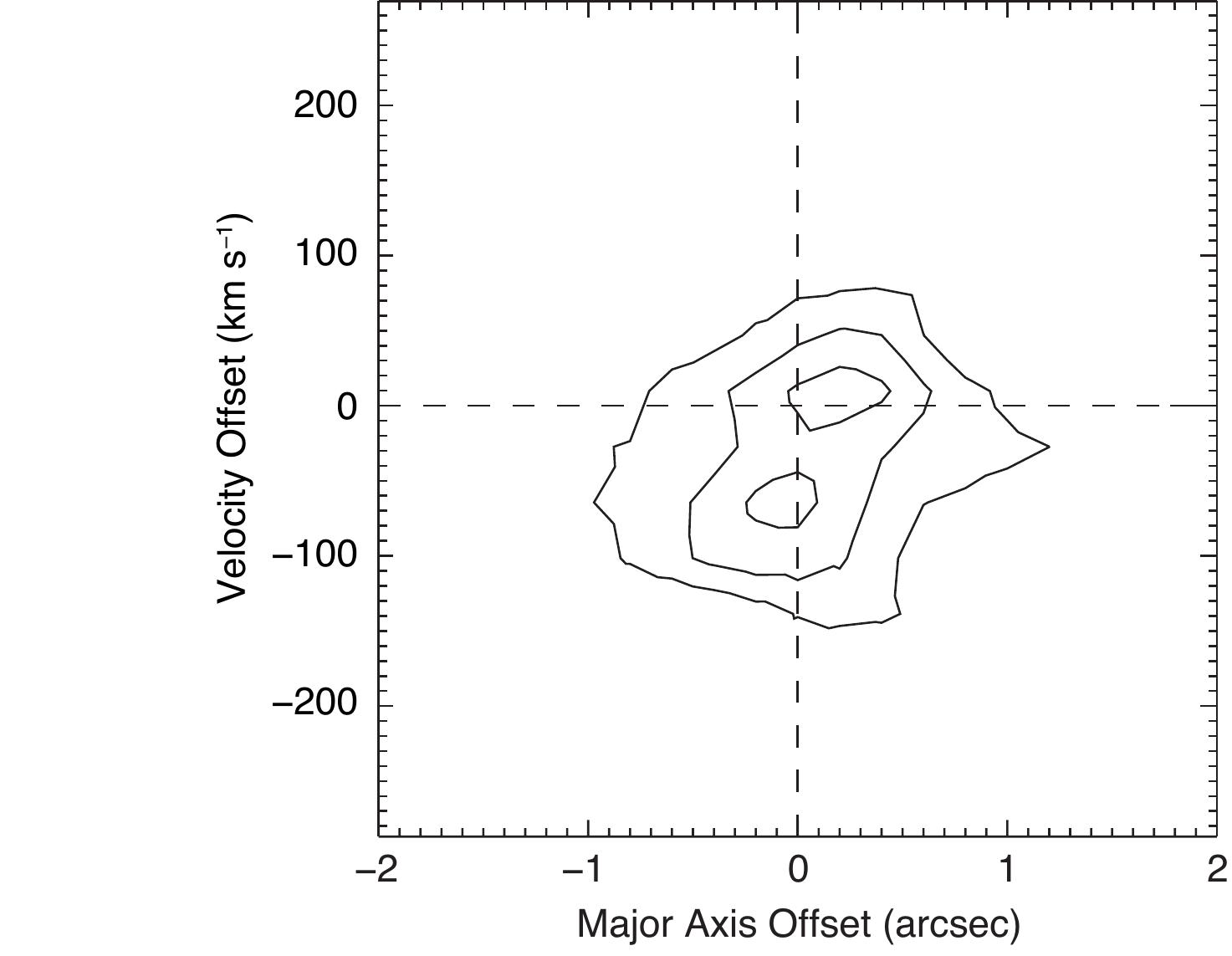}
\caption{Left/Center: Overlaid contours of two $10\,{\rm MHz}$ ($\sim110\,{\rm km\,s^{-1}}$) channels centered at $56.52\,{\rm km\,s^{-1}}$  (red lines) and $-54.75\,{\rm km\,s^{-1}}$ (blue lines) for the \mbox{CO(1--0)} line (left) and the ${\rm H\alpha}$ line (center). Contours are multiples of $\pm2\sigma$ ($\sigma_\text{1--0}=0.17\,{\rm mJy\,beam}^{-1}$; $\sigma_{\rm H\alpha}=1.18\times10^{-18}\,{\rm W\,\mu m^{-1}\,m^{-2}}$), where negative contours are dotted; synthesized beam/point spread function is shown at lower left. Right: Position-velocity diagram of the integrated line where the major axis (along the velocity gradient) is at a position angle of $75$ degrees that maximizes the velocity gradient. Contours are multiples of $1.75\,{\rm mJy\,beam}^{-1}$. \label{fig:renzo}}
\end{figure*}

The integrated intensity map (Fig.\,\ref{fig:intlinemap}) shows that the \mbox{CO(1--0)} emission is spatially extended in a northeast/southwest direction, parallel to the lensing shear \citep{smail2005}, and contains two peaks embedded within more extended emission, similar to those seen in ${\rm H\alpha}$ \citep{tecza2004}. The center of the \mbox{CO(1--0)} line emission is consistent with the peak positions of the \mbox{CO(3--2)} and \mbox{CO(7--6)} lines previously detected by \citet{downes2003} (Fig.\,\ref{fig:allmaps}), with $\sigma_\text{1--0}=0.^{\prime\prime}05$ based on an elliptical Gaussian fit in the image plane and the intrinsic uncertainty of the phase calibrator position. The Gaussian fit gives a major axis FWHM of $2.^{\prime\prime}6\pm0.^{\prime\prime}1$ at position angle $18.7\pm1.3$ degrees and a minor axis FWHM of $0^{\prime\prime}.8\pm0.^{\prime\prime}1$ when accounting for the beam size. As J14011 is lensed, these dimensions do not easily translate into rest-frame physical dimensions. While the line FWHM is quite narrow, the \mbox{CO(1--0)} emission does show a velocity gradient. In Fig.\,\ref{fig:renzo}, we plot overlaid channel maps of the red and blue halves of the central $200\,{\rm km\,s^{-1}}$ of \mbox{CO(1--0)} line as well as the same velocity ranges for the ${\rm H\alpha}$ integral field data presented in \citet{tecza2004}. The red and blue halves of both lines are slightly offset in similar directions, with the two peaks in the integrated line maps apparently dominated by emission at different velocities. In Fig.\,\ref{fig:renzo}, we also show a \mbox{CO(1--0)} position-velocity diagram for J14011 for a position angle of $75\pm5$ degrees that maximizes the velocity gradient.

We do not detect any significant continuum emission from J14011 in the (rest-frame) $2.6\,{\rm mm}$ continuum map produced from the line-free channels. We estimate a $3\sigma$ upper limit for the continuum emission of $26.7\,{\rm \mu Jy}$ for a point-like source, or $157\,{\rm \mu Jy}$ when scaled to the dimensions of the \mbox{CO(1--0)} emission.

In order to compare the \mbox{CO(1--0)} emission to existing optical and infrared data, we have taken care in aligning images. Based on the source sizes and structures in the two sets of maps in Fig.\,\ref{fig:renzo}, and the fact that ${\rm H\alpha}$ emission from H{\scshape ii} regions is expected to follow molecular gas at this linear resolution, we can use the positions of the northern and southern peaks to better align the datasets. The average offset between the peaks was applied as a shift to align the SINFONI ${\rm H\alpha}$ data cube with the radio observations. The measured offset is $0.^{\prime\prime}82$, which corresponds to roughly $\lesssim2\times$ the SINFONI point spread function (the astrometric uncertainty of the \mbox{CO(1--0)} map is $\sim0.^{\prime\prime}09$, including both statistical uncertainty and uncertainty due to the phase calibrator position). Based on this alignment, we use the continuum emission from the line-free channels of the SINFONI data-cube (at $\sim2.2\,\mu{\rm m}$) to extend the alignment to the {\it HST}/NICMOS F160W data of \citet{aguirre2012}. As we expect the continuum emission at both wavelengths to come from the same physical regions in the source, we use the continuum from the foreground interloper (J1c) and the UV-bright companion galaxy (J2) to determine the shift to apply to the {\it HST} image to align it with the ${\rm H\alpha}$ and thus \mbox{CO(1--0)} maps (Fig.\,\ref{fig:allmaps}). We measure an offset of $0.^{\prime\prime}49$ between the F160W and $2.2\,\mu{\rm m}$ maps, making the offset between the \mbox{CO(1--0)} and {\it HST} maps $<0.^{\prime\prime}5$.

\section{Analysis}
\label{sec:analysis}

\subsection{LVG modeling}
\label{sec:lvg}

J14011 has integrated brightness temperature ratios $r_{3,1}=0.97\pm0.16$ and $r_{7,1}=0.20\pm0.04$. Comparisons between the higher-$J$ and \mbox{CO(1--0)} emission (convolved to the resolutions of the higher-$J$ maps) do not show evidence for spatial variation of CO excitation. This value of $r_{3,1}$ is larger than the typical $r_{3,1}\sim0.6$ for SMGs and more similar to the $r_{3,1}\sim1$ seen in high-$z$ quasar host galaxies. The strong upper limits on X-ray emission and observations of mid-IR PAH features indicate that J14011 likely does not contain an AGN of sufficient strength to influence the global gas conditions, making its near-unity $r_{3,1}$ unusual for its class. However, among low-$z$ U/LIRGs (which also have an average $r_{3,1}\sim0.6$), there are a small number of starbursts that globally or locally have $r_{3,1}\sim1$ but lack AGN, such as NGC 3690 \citep{iono2009}, UGC 8739 \citep{yao2003}, the inner region of NGC 253 \citep[e.\/g.\/,][]{bayet2004}, and the northeast lobe of M82 \citep[e.\/g.\/,][]{ward2003}. While the high $r_{3,1}$ of J14011 is rare for a pure starburst, it is not unique. \citet{yao2003} propose that high CO excitation in U/LIRGs can be caused by a very high concentration of star formation, consistent with the finding of \citet{iono2009} that the peak central $r_{3,1}$ is greater than or equal to the global $r_{3,1}$ in spatially resolved maps. If the duration of highly concentrated star formation (such that the global and ``peak" CO excitation are identical) is short-lived, then it is sensible that only a small number of pure-starburst SMGs (or local U/LIRGs) will exhibit high $r_{3,1}$. Gas distributions that are compact relative to heating sources may be the  trait that such systems share with quasar host galaxies.

In order to investigate the physical conditions of the gas that produces these ratios, we have undertaken an excitation analysis using a Large Velocity Gradient (LVG) model. The details of the LVG modeling and analysis are described in Sharon et al. (in prep.). In short, the model follows the method of \citet{ward2002}, assumes a spherical cloud geometry, and is modified to include the effects of the cosmic microwave background (CMB). With observed CO transitions equal in number to the model parameters (kinetic temperature, ${\rm H_2}$ density, and CO column density per unity velocity gradient) and substantial uncertainties in the measured line ratios, we expect significant degeneracies between the allowed model parameters. We therefore use both a minimum $\chi^2$ and a Bayesian analysis \citep[the latter better capturing our uncertainties;][]{ward2003} to determine which model parameters best reproduce our measured line ratios. For the Bayesian analysis, we have applied logarithmic priors to the LVG model parameters to compensate for the large logarithmically-sampled parameter space, as opposed to the usual agnostic prior equal to unity.

The parameters of the minimum $\chi^2$ model are $T=136\,{\rm K}$, $n_{\rm H_2}=10^{3.2}\,{\rm cm^{-3}}$, and $N_{\rm CO}/\Delta v = 10^{17.25}\,{\rm cm^{-2}\,km^{-1}\,s}$.  However, there are significant degeneracies in all three parameters. For example, the Bayesian maximum probability model favors the opposite extreme in temperature-density parameter space: $T=24\,{\rm K}$, $n_{\rm H_2}=10^{5.6}\,{\rm cm^{-3}}$, and $N_{\rm CO}/\Delta v = 10^{16.75}\,{\rm cm^{-2}\,km^{-1}\,s}$. Similarly well-fitting models are produced for $20\,{\rm K}\lesssim T\lesssim150\,{\rm K}$ at comparable CO columns per velocity gradient ($10^{17\pm0.25}\,{\rm cm^{-2}\,km^{-1}\,s}$), where lower temperature models favor higher ${\rm H_2}$ densities ($\sim10^5\,{\rm cm^{-3}}$) and higher temperature models favor lower ${\rm H_2}$ densities ($\sim10^{3}\,{\rm cm^{-3}}$). The optical depths produced by these two models are very different from one another. The minimum $\chi^2$ model produces optically thick emission for the \mbox{CO(3--2)} and \mbox{CO(7--6)} transitions ($\tau_\text{3--2}=13.4$, $\tau_\text{7--6}=11.4$), and optically thin emission for the \mbox{CO(1--0)} transition ($\tau_\text{1--0}=0.7$). The cooler and more dense maximum probability model has the optical depths of the \mbox{CO(1--0)} and \mbox{CO(7--6)} lines reversed with $\tau_\text{1--0}=2.0$, $\tau_\text{3--2}=11.8$, and $\tau_\text{7--6}=0.7$. We also consider a restricted parameter space for the LVG models in which \emph{all} of the measured lines must be produced by optically thick emission. The minimum $\chi^2$ model with $\tau_{J_{\rm upper}\leq7}>1$ has a significantly lower temperature, $T=60\,{\rm K}$, $n_{\rm H_2}=10^{3.7}\,{\rm cm^{-3}}$, and $N_{\rm CO}/\Delta v = 10^{17.25}\,{\rm cm^{-2}\,km^{-1}\,s}$, while the maximum probability model only slightly increases in temperature (relative to the previous maximum probability model), with $T=28\,{\rm K}$, $n_{\rm H_2}=10^{5.0}\,{\rm cm^{-3}}$, and $N_{\rm CO}/\Delta v = 10^{16.75}\,{\rm cm^{-2}\,km^{-1}\,s}$. In these models, the optical depths of the \mbox{CO(1--0)} and \mbox{CO(3--2)} emission are comparable ($\tau_\text{1--0}\approx1.5$, $\tau_\text{3--2}\sim10$--$15$), but the \mbox{CO(7--6)} line is only just optically thick in the colder model, while in the warmer model the optical depth is a factor of $\sim10$ larger. The CO spectral line energy distributions (SLEDs) for these four models are shown in Fig.\,\ref{fig:lineratios}.

Our cooler and optically thick models favor kinetic temperatures comparable to the dust temperature, $T_{\rm dust}=42\,{\rm K}$ \citep{wu2009}, and are in line with the results of other LVG analyses of SMGs \citep[e.\/g.\/,][]{weiss2007,carilli2010,riechers2010b,danielson2011,riechers2011c}. Observations of different molecular or atomic species can be useful for breaking temperature degeneracies in CO LVG models. \citet{walter2011} make an independent estimate of the kinetic temperature in J14011 using their \mbox{C {\sc i}($^3 P_2\rightarrow {^3P_1}$)} line measurement in conjunction with the previously-measured \mbox{C {\sc i}($^3 P_1\rightarrow {^3P_0}$)} line {\citep{weiss2005a}, deriving $T_{\rm C\,{\scriptsize\textsc i}}=32.4\pm5.2\,{\rm K}$. At fixed $T_{\rm kin}=T_{\rm C\,{\scriptsize\textsc i}}$, the minimum $\chi^2$ model (which is indistinguishable from the maximum probability model due to similar probability and $\chi^2$ distributions in the $n_{\rm H_2}$--$N_{\rm CO}/\Delta v$ plane) has $n_{\rm H_2}=10^{4.6}\,{\rm cm^{-3}}$ and $N_{\rm CO}/\Delta v = 10^{17.00}\,{\rm cm^{-2}\,km^{-1}\,s}$ (Fig.\,\ref{fig:lineratios}) and produces optically thick emission for all three transitions ($\tau_\text{1--0}=2.1$, $\tau_\text{3--2}=16.0$, $\tau_\text{7--6}=2.7$). While a prior LVG analysis performed by \citet{bayet2009} preferred higher temperatures ($T=205\,{\rm K}$) and lower densities ($n_{\rm H_2}=10^{1.4}\,{\rm cm^{-3}}$) than any of our well-fitting models, due to the different goals of their work (predicting line strengths for a statistical sample rather than accurate determination of individual galaxy properties), given their use of only the \mbox{CO(3--2)} and \mbox{CO(7--6)} line measurements and our consistency with other estimates of the gas temperature, we view our results as more robust.

\begin{figure}
\plotone{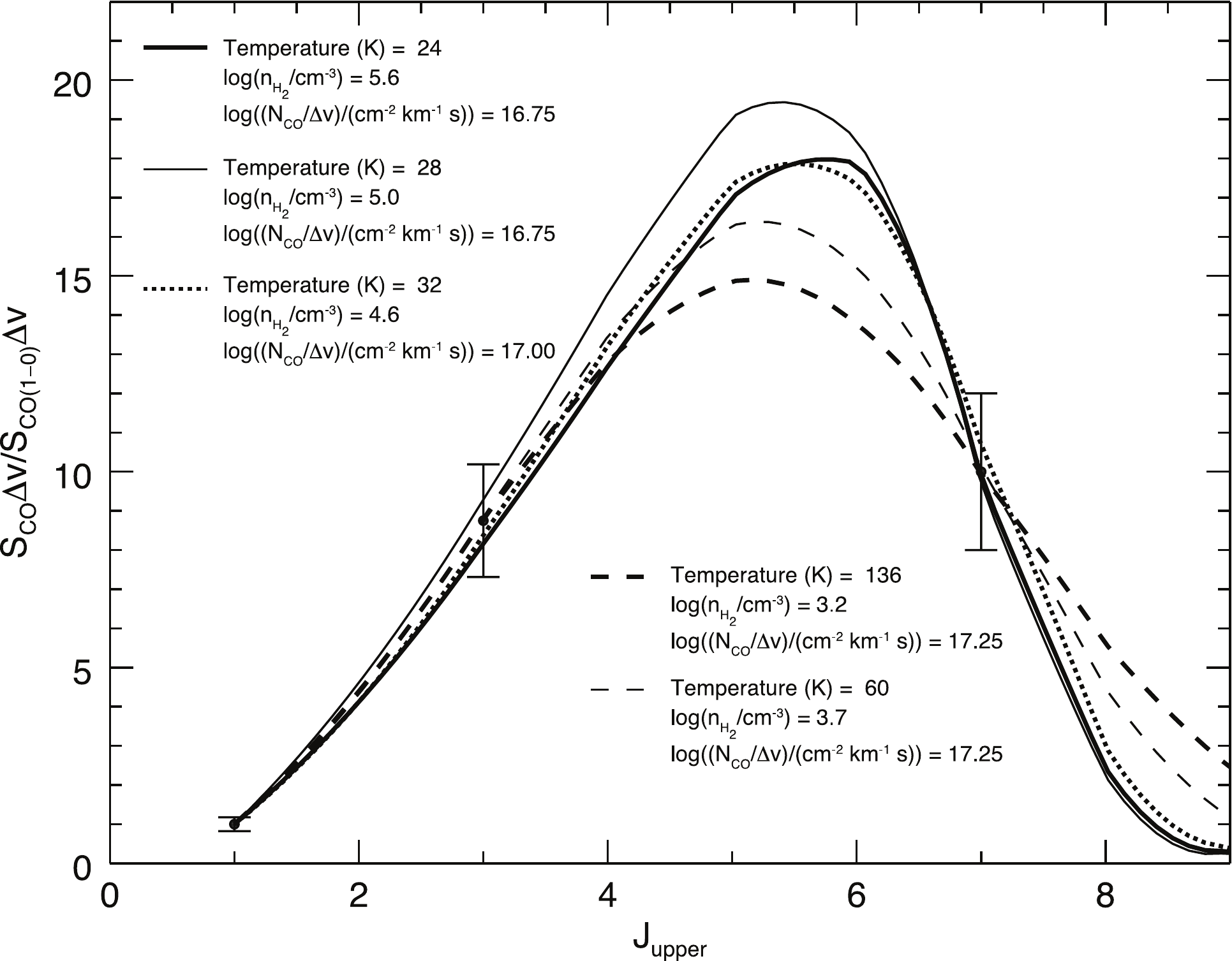}
\caption{Measured line ratios (points; in flux units) and CO SLEDs for the maximum probability model (thick solid line), maximum probability model requiring $\tau>1$ (thin solid line), minimum $\chi^2$ model (thick dashed line), minimum $\chi^2$ model with $\tau>1$ (thin dashed line), and the maximum probability/minimum $\chi^2$ model at fixed $T_{\rm C\,{\scriptsize\textsc i}}$ (thick dotted line). The LVG model parameters that produce these SLEDs are as labeled. \label{fig:lineratios}}
\end{figure}

Besides the degeneracies between the best-fit models noted above, it is worth considering what degeneracies exist \emph{within} each model (e.\/g.\/, degeneracies between $n_{\rm H_2}$ and $N_{\rm CO}/\Delta v$ for a specific choice of temperature, or between $T_{\rm kin}$ and $n_{\rm H_2}$ for a specific value of $N_{\rm CO}/\Delta v$). In Fig.\,\ref{fig:contours_nco} we have plotted curves of constant probability (or $\chi^2$) on the temperature-density plane for the best-fit values of $N_{\rm CO}/\Delta v$. For a specific value of $N_{\rm CO}/\Delta v$, there is generally an ``L"-shaped region with nearly constant probability/$\chi^2$ that will reasonably reproduce a given CO line ratio. With multiple line ratios, the locus where the two ``L"-shaped regions overlap produces the best-fit models. For multiple line ratios, these ``L"-shaped regions overlap more for emission lines with similar optical depth ratios. The $\tau>1$ regions are marked with dashed lines in Fig.\,\ref{fig:contours_nco} and are also ``L"-shaped. As optical depth is proportional to $N_{\rm CO}/\Delta v$, larger regions satisfy the $\tau>1$ criterion for larger values of $N_{\rm CO}/\Delta v$.

\begin{figure*}
\plottwo{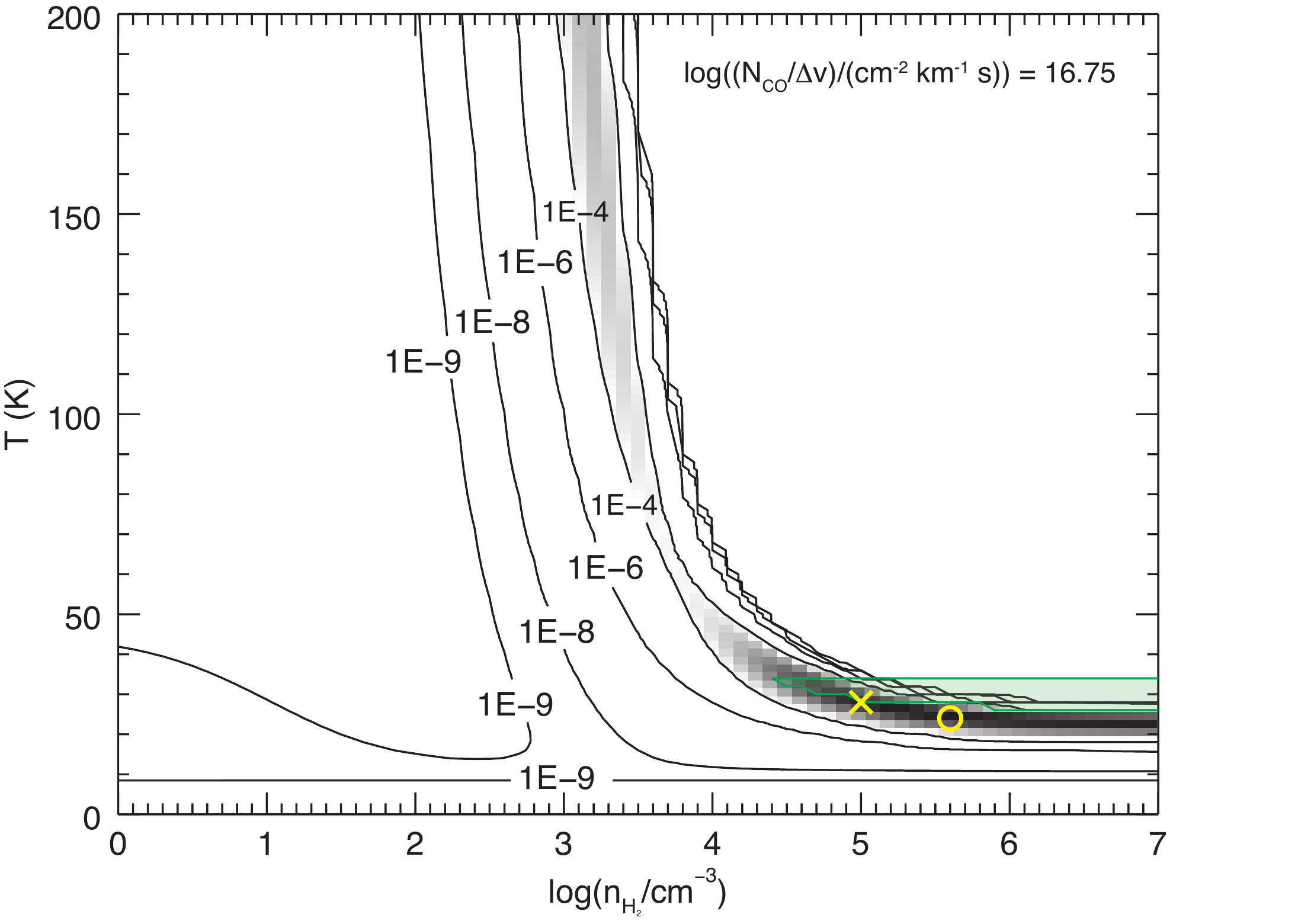}{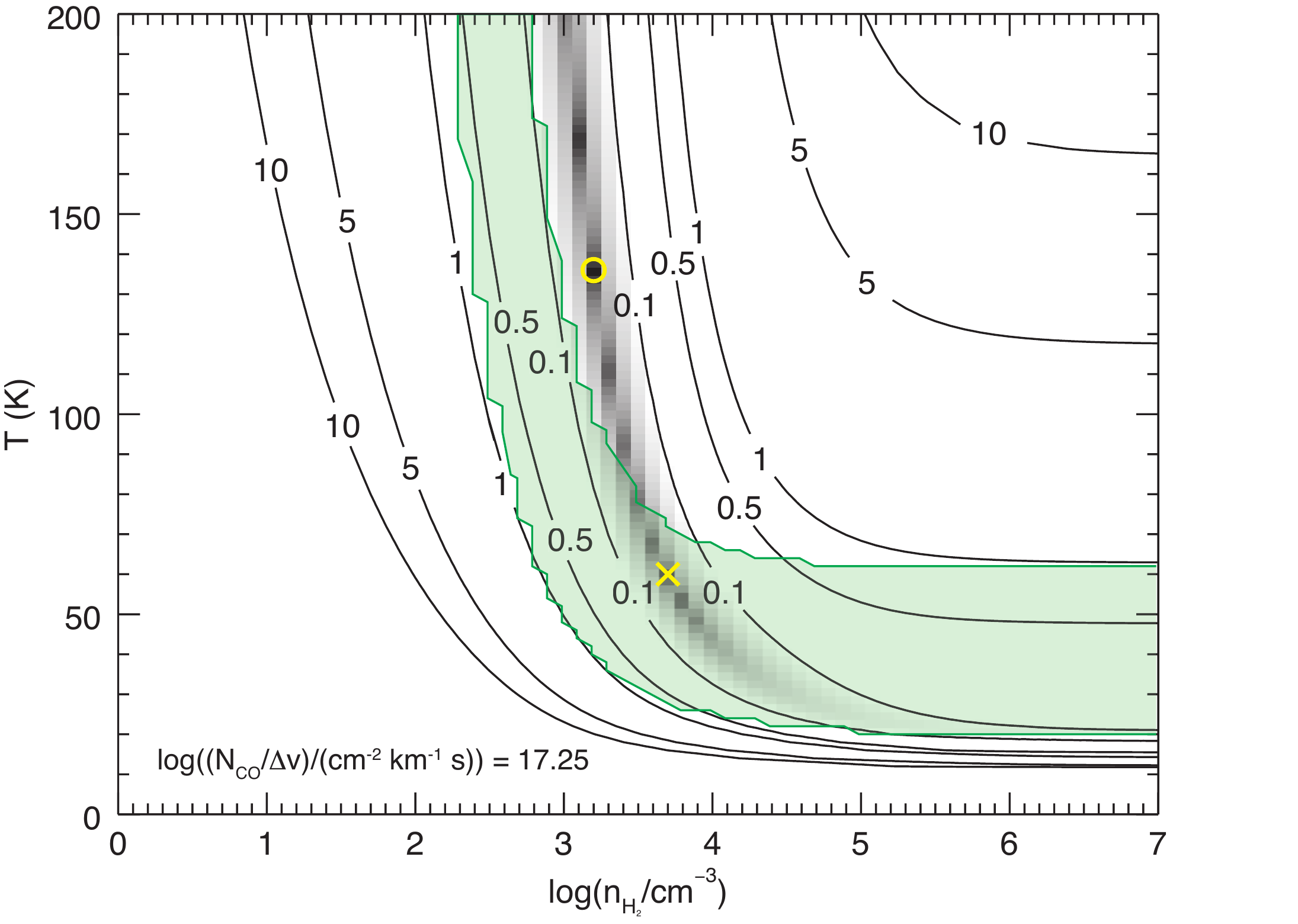}
\caption{Contours of constant probability (left) and $\chi^2$ (right) in the temperature-density plane for the best-fit values of $N_{\rm CO}/\Delta v$. Models with higher probability than the highest contour (or lower $\chi^2$ than the lowest contour) have been logarithmically shaded in greyscale. Models with parameters outside of the green-shaded region produce optically thin emission in at least one line. The best-fit models with and without the optical depth cut have been marked with yellow crosses and circles respectively. \label{fig:contours_nco}}
\end{figure*}

For a specific temperature, there are also degeneracies between the ${\rm H_2}$ density and CO column density per velocity gradient. In Fig.\,\ref{fig:contours_t} we have plotted curves of constant $\chi^2$ in the $n_{\rm H_2}$--$N_{\rm CO}/\Delta v$ plane for the optically thick model temperatures that give highest probability and minimum $\chi^2$ (in this plane, the probability and $\chi^2$ contours have the same shape, but are scaled differently since the probability is $\propto e^{-\chi^2}$). While slightly shifted between the panels for the two temperatures, there is a clear linear valley with low $\chi^2$. This region corresponds to $N_{\rm CO}/\Delta v\propto n_{\rm H_2}^{-1}$, and it exists for all $T_{\rm kin}>30\,{\rm K}$ models (for $T<30\,{\rm K}$, the valley of low $\chi^2$ values asymptotes to a single value of  $N_{\rm CO}/\Delta v$ for higher ${\rm H_2}$ densities). This linear feature corresponds to a constant optical depth ratio for the emission lines that best reproduce the measured line ratios.

\begin{figure*}
\plottwo{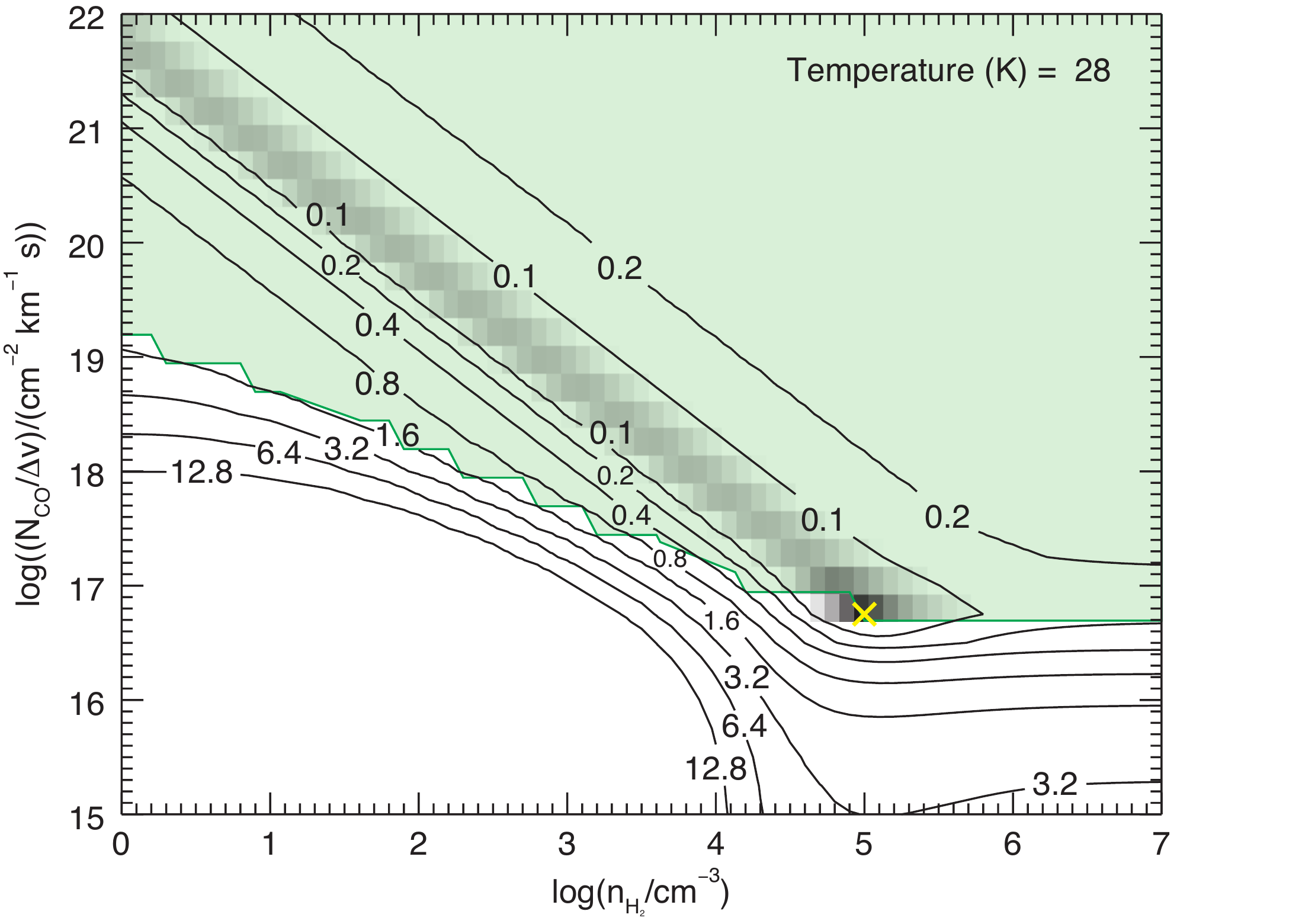}{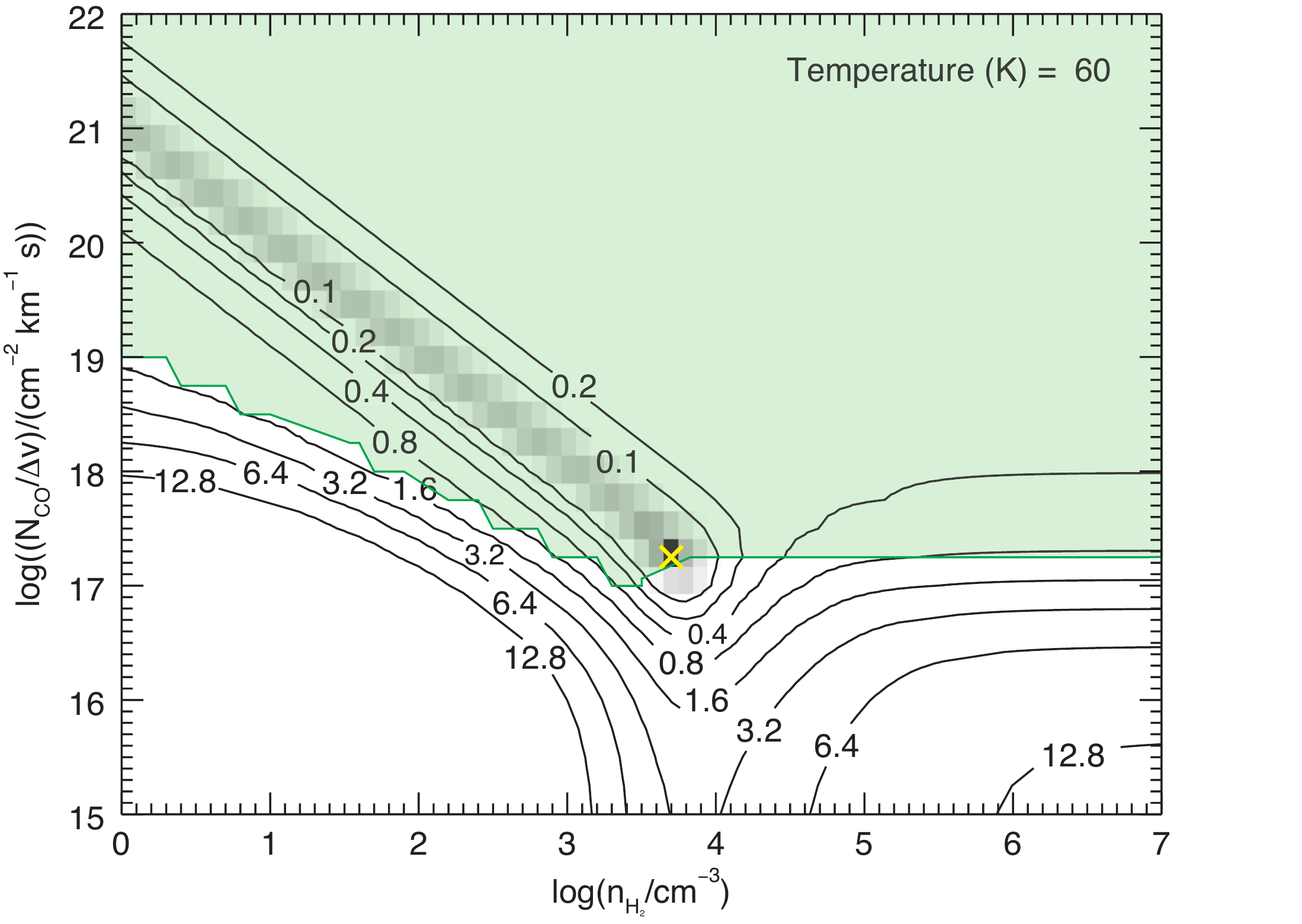}
\caption{Contours of constant $\chi^2$ in the $n_{\rm H_2}$--$N_{\rm CO}/\Delta v$ plane for highest probability optically thick model ($28\,{\rm K}$; left) and minimum $\chi^2$ optically thick model ($60\,{\rm K}$; right). Models with $\chi^2<0.1$ have been logarithmically shaded in greyscale. Models with parameters that fall below the green-shaded region produce optically thin emission. The highest probability/minimum $\chi^2$ models in each plane are marked by the yellow crosses. \label{fig:contours_t}}
\end{figure*}

With these substantial degeneracies it is worth considering priors in addition to optical depth that might limit the allowed parameter ranges for the LVG models. Since the most dense molecular gas in the Galaxy is found in star-forming virialized clouds, we can consider placing limits on the density by considering the degree of virialization as parameterized by

\begin{equation}
K_{\rm vir} \sim 1.54\frac{[{\rm CO/H_2}]}{\sqrt{a} \Lambda_{\rm CO}}\left[ \frac{\langle n_{\rm H_2}\rangle}{10^3\,{\rm cm^{-3}}}\right]^{-1/2}
\label{eq:alpha}
\end{equation}

\noindent
where $\Lambda_{\rm CO}$ is the CO abundance per unit velocity gradient and $a\sim1$--$2.5$ is a constant that depends on the cloud geometry \citep[e.\/g.\/,][]{goldsmith2001,greve2009}. In this parameterization, ${\rm [CO/H_2]}\,\Lambda_{\rm CO}^{-1}$ is the velocity gradient in the molecular gas, and $K_{\rm vir}\ll1$ is dynamically unobtainable. For J14011, the velocity gradient could be a value determined by global parameters (given by the $\sim150\,{\rm km\,s^{-1}}$ \mbox{CO(1--0)} FWHM and approximate delensed optical source size of $\sim10\,{\rm kpc}$ from \citet{smail2005}) if the line luminosity is tracing the total gravitational potential of the galaxy \citep[e.\/g.\/,][]{downes1993}, or a different value if the molecular gas is in individual star-forming clouds. For the global velocity gradient, requiring that $K_{\rm vir}> 0.1$ means that $n_{\rm H_2}>10^{1.5}\,{\rm cm^{-3}}$ is not allowed, excluding all of the well-fitting models discussed above. In general, low $\chi^2$ models with low densities that meet the $K_{\rm vir}> 0.1$ criterion under-predict the \mbox{CO(3--2)/CO(1--0)} line ratio by $\gtrsim1\sigma_{3,1}$. Velocity gradients larger than the global value would allow the higher densities of our best-fit LVG models to satisfy the virialization requirement. In order to allow densities as high as $10^4\,{\rm cm^{-3}}$ ($10^5\,{\rm cm^{-3}}$) with $K_{\rm vir}>0.1$, the velocity gradient would need to be $>0.27\,{\rm km\,s^{-1}\,pc^{-1}}$ ($>0.86\,{\rm km\,s^{-1}\,pc^{-1}}$.) If the individual clouds are virialized ($K_{\rm vir}=1$), these ${\rm H_2}$ densities require velocity gradients of $\sim3\,{\rm km\,s^{-1}\,pc^{-1}}$ ($\sim9\,{\rm km\,s^{-1}\,pc^{-1}}$).

Models with low $n_{\rm H_2}$ and high $N_{\rm CO}/\Delta v$ can be eliminated by requiring the CO column length to be less than the maximum source size. However, this prior requires the assumption of an upper limit for $[{\rm CO/H_2}]$. Conservative limits (i.\/e.\/, $[{\rm CO/H_2}]<5\times10^{-4}$; \citealt{ward2003}) exclude clouds with densities comparable to or less than that seen in the Milky Way ($\sim10^2\,{\rm cm^{-3}}$) along the low-$\chi^2$ degeneracy valley in the $n_{\rm H_2}$--$N_{\rm CO}/\Delta v$ plane. While this prior is not useful for distinguishing among the best-fit models for J14011, it does support the higher ${\rm H_2}$ densities preferred by the LVG analysis.

While it is unlikely that a simple LVG model will accurately capture the full range of temperatures and densities that exist in real molecular clouds, considering the degeneracies discussed above, we conclude that the bulk of the molecular gas likely has, on average, $T_{\rm kin}=20$--$60\,{\rm K}$, $n_{\rm H_2}\sim10^4$--$10^5\,{\rm cm^{-3}}$, and $N_{\rm CO}/\Delta v=10^{17.00\pm0.25}\,{\rm cm^{-2}\,km^{-1}\,s}$. The maximum likelihood for the thermal pressure peaks at $\sim10^2\,{\rm K\,cm^{-3}}$, but has a significant tail up to $10^6\,{\rm K\,cm^{-3}}$. Measurements of other mid-$J$ CO lines, like \mbox{CO(5--4)}, would help to better constrain these models. Using our best estimates for the LVG model parameters, we constrain $[{\rm CO/H_2}]/(\Delta v/\Delta r)=10^{-6.7}$--$10^{-5.2}\,{\rm pc\,km^{-1}\,s}$, which is close to the standard assumption of $10^{-5}\,{\rm pc\,km^{-1}\,s}$ used in other LVG analyses. While the measured line ratios in J14011 appear consistent with a single phase molecular ISM, analysis of other SMGs \cite[e.\/g.\/,][]{harris2010,ivison2011,danielson2011,bothwell2012} and observations of local galaxies \citep[e.\/g.\/,][]{yao2003} indicate that gas in star-forming galaxies is multi-phase; i.\/e.\/, it is unlikely that the gas responsible for the bulk of the \mbox{CO(1--0)} emission is the same gas that is emitting the \mbox{CO(7--6)} line, which is effectively what we are requiring by considering only single-phase models. With only three measured lines, the results of a multi-phase LVG analysis would be far too degenerate to draw meaningful conclusions about the state of molecular gas. Higher resolution observations would be helpful in establishing whether the CO lines are or are not being emitted from the same physical volumes of gas. Based on the agreement of the global line profiles for the existing CO measurements it is unlikely that excitation differences exist on large scales across J14011, but local variations with broadly similar kinematics could exist on scales below our current resolution limit.

Based on the CO abundance per unit velocity gradient of the best-fit LVG models, we can determine the CO-to-${\rm H_2}$ abundance for an assumed velocity gradient. Using the global velocity gradient from our measured line width and the \citet{smail2005} source size, we estimate the CO-to-${\rm H_2}$ abundance to be $[{\rm CO/H_2}]=10^{-8.6}$--$10^{-7.1}$, which is very low. Alternatively, if we assume more standard values of $[{\rm CO/H_2}]=10^{-5}$--$10^{-4}$, the best-fit model densities return higher velocity gradients of $\sim1$--$30\,{\rm km\,s^{-1}\,pc^{-1}}$, a range that includes the necessary gradients for our best-fit models to meet the virialization criteria.

Since LVG modeling produces reasonable physical conditions for J14011, and it is only the introduction of the global velocity gradient that produces very low CO-to-${\rm H_2}$ abundance and unphysical virialization parameters, we conclude that J14011 likely has large local velocity gradients. Higher velocity gradients are possible if J14011 has a face-on orientation and contains numerous small clouds below our resolution limit that have turbulent velocities and/or bulk motions summing to the measured ${\rm FWHM_\text{1--0}}\approx150\,{\rm km\,s^{-1}}$. If the individual clouds are virialized, their velocity gradients are large enough to produce more reasonable values of the CO-to-${\rm H_2}$ abundance for our best-fit densities. Although such velocity gradients would cause J14011's molecular clouds to fall above the Galactic line-width vs.~size relation \citep{larson1981}, such behavior has been seen in another $z\sim2$ SMG, SMM\,J2135-0102, which has a mean internal velocity gradient of $\sim0.5\,{\rm km\,s^{-1}\,pc^{-1}}$ for individual clumps \citep{swinbank2011}. \citet{swinbank2011} also find similarly high densities, and thus infer large velocity gradients under the assumption of virialization, for the molecular gas in SMM\,J2135-0102.

\subsection{Resolved star formation law}

Using the aligned ${\rm H\alpha}$ and \mbox{CO(1--0)} maps, we plot a spatially resolved Schmidt-Kennicutt relation \citep{schmidt1959,kennicutt1998} for J14011 in Fig.\,\ref{fig:kslaw}. We have smoothed the integrated line maps to the same spatial resolution ($0.^{\prime\prime}70\times0.^{\prime\prime}65$), resampled to the pixel size from the SINFONI data ($0.^{\prime\prime}25$), and excluded pixels that do not have at least $2\sigma$ significance in both maps. In order to convert the ${\rm H\alpha}$ surface luminosity to a star formation rate surface density, we use the $L_{\rm H\alpha}$-SFR conversion factor given in \citet{kennicutt1998} scaled to match the \citet{calzetti2007} initial mass function (IMF) used in \citet{bigiel2008,bigiel2010}. We have also applied an additional correction to reflect the fact that the observed ${\rm H\alpha}$ luminosity underestimates the galaxy's total star formation rate: based on the \cite{kennicutt1998} calibration rescaled to the \citet{calzetti2007} IMF, J14011's total FIR-determined SFR is $400\,M_\odot\,{\rm yr^{-1}}$, exceeding its ${\rm H\alpha}$-determined SFR by a factor of fifteen. Given that J14011 does not contain a strong AGN \citep{rigby2008}, that there is good agreement between the CO and the ${\rm H\alpha}$ morphologies (unlikely if the FIR vs.~${\rm H\alpha}$ discrepancy were determined by {\it large-scale} patchiness in obscuration), and that the correction for the Balmer decrement as in \citet{tecza2004} would already account for a factor of 1.5 (of the factor of 15 difference), we have simply scaled the ${\rm H\alpha}$ surface brightness of each pixel by the {\it global} ratio ${\rm SFR_{FIR}/SFR_{H\alpha}}$ before plotting in Fig.\,\ref{fig:kslaw}.

\begin{figure*}
\plotone{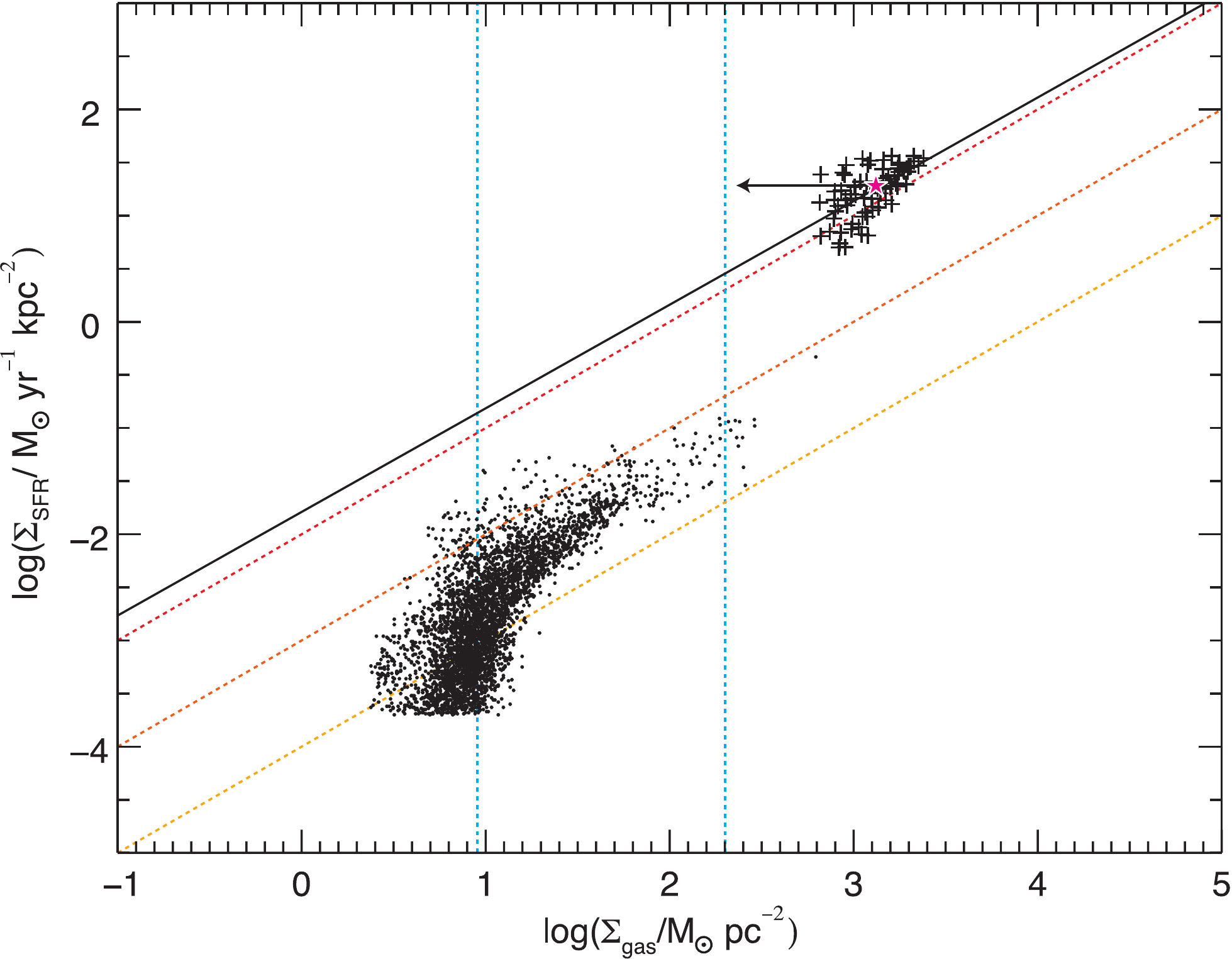}
\caption{Star formation rate surface density as measured by ${\rm H\alpha}$ surface brightness (rescaled by the global ${\rm SFR_{FIR}/SFR_{H\alpha}}$) vs.~molecular gas mass surface density for J14011 (crosses) and total gas mass surface density for local disk galaxies from \citet{bigiel2010} (dots; clipped at $2\times10^{-4}\,M_\sun\,{\rm yr^{-1}\,kpc^{-2}}$). The average value for J14011 is given by the star, and a power law fit to the resolved surface density points is given by the solid line. The arrow pointing to the left shows where J14011's locus of points (centered at the average value) would move if the molecular gas mass were calculated using the ``starburst" value of $\alpha_{\rm CO}$ instead of the ``disk" value. Dotted lines are the same as in \citet{bigiel2008}, where the diagonal lines represent the constant star formation efficiencies required to consume $1\%$ (yellow), $10\%$ (orange), and $100\%$ (red) of the gas in $10^8$ years, the left vertical line shows the H{\scshape i} surface density saturation, and the right vertical line marks the proposed transition between star formation laws at $200\,M_\sun\,{\rm pc^{-2}}$. \label{fig:kslaw}}
\end{figure*}

Converting \mbox{CO(1--0)} luminosity to molecular gas mass requires the assumption of a conversion factor $\alpha_{\rm CO}$. Based on the results of our LVG analysis above, which produces more self-consistent molecular gas properties if the molecular ISM in J14011 is composed of multiple unresolved clouds rather than a single large gas reservoir, we use a fiducial Milky Way disk value of $\alpha_{\rm CO}=4.6\,M_\sun\,{\rm (K\,km\,s^{-1}\,pc^2)}^{-1}$ \citep{solomon1991}. This choice is consistent with the proposed continuous form of $\alpha_{\rm CO}$ of \citet{narayanan2012} for the measured metallicity of component J1 from \citet{tecza2004} of $12+{\rm log(O/H})=8.96^{+0.06}_{-0.10}$. In Fig.\,\ref{fig:kslaw}, the gas mass surface densities for both J14011 and the low-redshift galaxies have been scaled to account for helium. Since gravitational lensing conserves surface brightness, magnification corrections are to first order unnecessary for evaluating the resolved Schmidt-Kennicutt relation for J14011.

The resolved star formation law for J14011, as probed by the FIR-corrected ${\rm H\alpha}$ map, is not consistent with that of the local disk galaxies from \citet{bigiel2008} despite our use of a Milky Way gas mass conversion factor. While \citet{bigiel2008} use a sub-Milky Way value of $\alpha_{\rm CO}=3.2\,M_\sun\,{\rm (K\,km\,s^{-1}\,pc^2)}^{-1}$, any value of $\alpha_{\rm CO}<4.6$ moves J14011 further above the local relation, and $\alpha_{\rm CO}=0.8$ moves it far above. Our result supports the conclusions of \citet{genzel2010}, indicating that SMGs' position above the local star formation relation is not simply due to the choice of CO-to-${\rm H_2}$ conversion factor. Correcting for different IMFs while respecting the differing choices in $\alpha_{\rm CO}$, the star formation law for J14011 is consistent with both the SMGs in \citet{genzel2010} (who use $\alpha_{\rm CO}=1.0$) and the local starbursts from \citet{kennicutt1998} (who uses $\alpha_{\rm CO}=0.8$), and is not consistent with the high-$z$ ``normal" star-forming galaxies from \citet{daddi2010} (who use $\alpha_{\rm CO}=3.6$). Rescaling J14011 to the lower $\alpha_{\rm CO}$ used in \citet{genzel2010} or \citet{kennicutt1998} would move J14011 far above the star formation relation for SMGs and low-$z$ starbursts, further supporting our choice of a Galactic CO-to-${\rm H_2}$ conversion factor for this system.

To allow comparisons to other work, we fit the star formation and molecular gas surface densities to a power law and formally retrieve a near-unity index of $0.98\pm0.14$ (albeit with $\chi^2=2.48$), consistent with indices measured in local spirals \citep{bigiel2008}. Our application of the scaling factor to account for dust-obscured star formation does not affect our measured index, as appropriate if ${\rm H\alpha}$ traces the location of J14011's star-forming regions, even if it does not capture the total SFR. We note that in our pixel-to-pixel comparison, the data points are not all independent due to our oversampling of the beam/point spread function; the points in Fig.\,\ref{fig:kslaw} are drawn from a total area of $\sim10$ resolution elements.

\section{Conclusions}
\label{sec:concl}

We present high-resolution \mbox{CO(1--0)} mapping of the submillimeter galaxy SMM\,J14011+0252. Based on a spectrally-resolved detection, we find a revised near-unity value for the \mbox{CO(3--2)}/ \mbox{CO(1--0)} line ratio ($r_{3,1}=0.97\pm0.16$). Although  $r_{3,1}$ is closer to the average value seen in quasar-host galaxies than in SMGs, given that similar values are seen in some low-$z$ ULIRGs that lack AGN, $r_{3,1}\sim1$ is not inconsistent with previous observational limits on the AGN strength in J14011. J14011 may be experiencing a short period of highly-concentrated star formation affecting the gas excitation \citep[e.\/g.\/,][]{yao2003,iono2009}, indicating that $r_{3,1}\sim1$ is not a perfect discriminant between starbursts and quasar host galaxies at high redshift. We find that the position of the \mbox{CO(1--0)} emission agrees with those of previously detected mid-$J$ lines \citep{downes2003}. The \mbox{CO(1--0)} emission is spatially extended from the southeast to northwest, extending between the J1s and J1n components previously identified at optical wavelengths, and showing two peaks that are slightly offset in velocity. A face-on interpretation is favored by our LVG models and is consistent with J14011's narrow line width.

Although the results of the LVG modeling are highly degenerate and we are restricted to a single-phase analysis, our determination of the gas kinetic temperature is consistent with the gas temperature measured using C {\sc i} \citep{walter2011} and with the dust temperature \citep{wu2009}. Our best-fit models give $n_{\rm H_2}\sim10^4$--$10^5\,{\rm cm^{-3}}$ and $N_{\rm CO}/\Delta v=10^{17.00\pm0.25}\,{\rm cm^{-2}\,km^{-1}\,s}$. These densities only produce reasonable [CO/${\rm H_2}$] values and satisfy dynamical constraints if the velocity gradient is larger than the galaxy wide-average, implying that J14011's molecular ISM comprises numerous unresolved molecular clouds.

Based on the agreement between the metallicity-dependent version of the CO-to-${\rm H_2}$ conversion factor from \citet{narayanan2012} and the canonical ``disk" value favored by our LVG analysis, we use $\alpha_{\rm CO}=4.6\,M_\sun\,{\rm (K\, km\,s^{-1}\,pc^2)^{-1}}$ to derive a molecular gas mass of $(1.9\pm0.3)\times10^{11}\,M_\sun$ (corrected for $\mu=3.5$; \citealt{smail2005}). Using ${\rm H\alpha}$ data from \citet{tecza2004}, corrected by the SFR measured in the FIR, we are able to compare the spatially-resolved Schmidt-Kennicutt relation in J14011 to those of other galaxies. We find that J14011 falls above the star formation rate vs.~molecular gas surface density relation seen in normal star-forming galaxies \citep[e.\/g.\/,][]{bigiel2008}, even when we use the local ``disk" value of the CO-to-${\rm H_2}$ conversion factor. This result indicates that the observed offset of starburst galaxies from the local universe Schmidt-Kennicutt law is not solely due to the use of two different gas mass conversion factors, in agreement with \citet{genzel2010}. J14011 is consistent in its star formation relation with other SMGs and starbursts when we respect different authors' choices of gas mass conversion factor. The near-unity index of the power law fit to J14011's points is comparable to indices measured in the local universe.

\acknowledgments{We thank an anonymous referee for helpful comments. We also thank Rob Ivison for his help obtaining the \mbox{CO(1--0)} observations and for useful comments, Claire Chandler for sharing insights on calibration, Ric Davies for providing the ${\rm H\alpha}$ channel maps, and Dennis Downes for sharing the \mbox{CO(3--2)} and \mbox{CO(7--6)} data. This work has been supported by NSF grant AST-0708653 and NASA grant HST-GO-11143.01-A. CES has received support from an AAUW American Fellowship.}

\end{document}